\documentclass[aps,prl,twocolumn,showpacs,superscriptaddress,floatfix,longbibliography]{revtex4-1}

\usepackage{graphicx}
\usepackage{dcolumn}
\usepackage{bm}
\usepackage{amssymb}
\usepackage{microtype}
\usepackage{xfrac}
\usepackage{gensymb}
\usepackage{xcolor}
\usepackage{array}
\newcolumntype{C}[1]{>{\centering\arraybackslash}p{#1}}

\begin{document}

\title{Complex Transport and Magnetism in Inhomogeneous Mixed Valence Ce$_3$Ir$_4$Ge$_{13}$}

\author{A.~M.~Hallas}
\altaffiliation[Current affiliation: ]{Department of Physics \& Astronomy and Quantum Matter Institute, University of British Columbia}
\email[Email: ]{alannah.hallas@ubc.ca}

\author{C.~L.~Huang}
\affiliation{Department of Physics and Astronomy and Rice Center for Quantum Materials, Rice University, Houston, TX, 77005 USA}

\author{Binod~K.~Rai}
\affiliation{Department of Physics and Astronomy and Rice Center for Quantum Materials, Rice University, Houston, TX, 77005 USA}

\author{A.~Weiland}
\affiliation{Department of Chemistry \& Biochemistry, University of Texas at Dallas,
Richardson, Texas 75080, United States}

\author{G.~T.~McCandless}
\affiliation{Department of Chemistry \& Biochemistry, University of Texas at Dallas,
Richardson, Texas 75080, United States}

\author{Julia~Y.~Chan}
\affiliation{Department of Chemistry \& Biochemistry, University of Texas at Dallas,
Richardson, Texas 75080, United States}

\author{J.~Beare}
\affiliation{Department of Physics and Astronomy, McMaster University, Hamilton, Ontario L8S 4M1, Canada}

\author{G.~M.~Luke}
\affiliation{Department of Physics and Astronomy, McMaster University, Hamilton, Ontario L8S 4M1, Canada}

\author{E.~Morosan}
\affiliation{Department of Physics and Astronomy and Rice Center for Quantum Materials, Rice University, Houston, TX, 77005 USA}
\email[Email: ]{emorosan@rice.edu}

\date{\today}

\begin{abstract}

We report the discovery of Ce$_3$Ir$_4$Ge$_{13}$, a new Remeika phase compound with a complex array of structural, electronic, and magnetic properties. Our single crystal x-ray diffraction measurements show that Ce$_3$Ir$_4$Ge$_{13}$ forms in the tetragonally distorted $I4_1/amd$ space group. The electrical resistivity is almost temperature independent over three decades in temperature, from 0.4 K to 400 K, while the Hall coefficient measurements are consistent with a low-carrier semimetal. Magnetic susceptibility measurements reveal an effective moment of $\mu^{\text{exp}}_{\text{eff}} = 1.87~\mu_B$/Ce, suggesting that this material has a mixture of magnetic Ce$^{3+}$ and non-magnetic Ce$^{4+}$. Upon cooling, Ce$_3$Ir$_4$Ge$_{13}$ first enters a short range magnetically ordered state below $T_{\text{SRO}}=10$ K, marked by a deviation from Curie-Weiss behavior in susceptibility and a broad field-independent heat capacity anomaly. At lower temperatures, we observe a second, sharper peak in the heat capacity at $T^* = 1.7$ K, concurrent with a splitting of the field-cooled and zero-field-cooled susceptibilities. A small resistivity drop at $T^*$ suggests a loss of spin disorder scattering consistent with a magnetic ordering or spin freezing transition. Ce$_3$Ir$_4$Ge$_{13}$ is therefore a rare example of an inhomogeneous mixed valence compound with a complex array of thermodynamic and transport properties.

\end{abstract}

\maketitle

\section{Introduction}

Remeika phase materials, which have the chemical formula $R_3T_4M_{13}$ ($R =$~rare earth, $T =$~transition metal, $M = p-$block metal), are a chemically versatile family of intermetallics. This structure type, which is a derivative of the simple perovskite structure, shares cage-like structural motifs with clathrate and filled-skutturdite materials. The ideal Remeika phase (cubic space group $Pm\bar{3}n$), which is also known as the Yb$_3$Rh$_4$Sn$_{13}$ structure type, was first identified in 1980~\cite{remeika1980new,hodeau1980crystal,cooper1980x,vandenberg1980crystallography} and more than 120 materials are now known to form in this structure type and its seven lower symmetry derivatives, which include tetragonal, monoclinic, and trigonal distortions~\cite{oswald2017proof,gumeniuk2018structural}. Remeika phase materials exhibit a complex interplay between their lattice, charge, and spin degrees of freedom, resulting in an array of physical properties, including superconductivity, structural phase transitions, heavy fermion behavior, and quantum criticality~\cite{gumeniuk2018structural}.    

The Ce-based subset of Remeika compounds, Ce$_3T_4M_{13}$, is particularly noteworthy. Materials in this family range from non-magnetic metals (\emph{e.g.} Ce$_3$Rh$_4$Pb$_{13}$~\cite{sokolov2007crystalline}) to antiferromagnetically ordered heavy fermion compounds (\emph{e.g.} Ce$_3$Ir$_4$Sn$_{13}$~\cite{sato1993magnetic}). Two members of this family, Ce$_3$Co$_4$Sn$_{13}$~\cite{otomo2016chiral} and Ce$_3$Rh$_4$Sn$_{13}$~\cite{suyama2018chiral}, have been found to undergo a subtle structural phase transition from the ideal $Pm\bar{3}n$ phase to a chiral structure in space group $I2_13$ at $T_D=160$~K and 350~K, respectively. Other noteworthy members of this family are inhomogeneous mixed valence Ce$_3$Ru$_4$Ge$_{13}$~\cite{ghosh1995crystal} and Ce$_3$Os$_4$Ge$_{13}$~\cite{prakash2016ferromagnetic}, where significant site-mixing between Ce and Ge results in a mixture of magnetic Ce$^{3+}$ and non-magnetic Ce$^{4+}$ ($\alpha$-like and $\gamma$-like, respectively, in analogy to elemental cerium~\cite{koskenmaki1978cerium}).

An open question that relates to many members of the Ce$_3T_4M_{13}$ family is the nature of their magnetic ground states. A large subset of these compounds, including Ce$_3$Pt$_4$In$_{13}$~\cite{hundley2001unusual}, Ce$_3$Co$_4$Sn$_{13}$~\cite{cornelius2006observation,slebarski2012electronic}, and Ce$_3$Rh$_4$Sn$_{13}$~\cite{kohler2007low,oduchi2007magnetic,slebarski2012electronic}, exhibit broad peaks in their magnetic heat capacity near 1~K, coincident with a small drop in resistivity. The entropy release associated with these heat capacity anomalies is close to the expected $R\ln{2}$ for an isolated ground state doublet. Consequently, several works have interpreted these heat capacity peaks as originating from the onset of long-range magnetic order~\cite{hundley2001unusual,oduchi2007magnetic}. However, in cases where neutron diffraction measurements have been performed, no magnetic Bragg peaks are observed~\cite{christianson2007low,suyama2018chiral}. Thus, while it is clear that coherent spin fluctuations develop in these Ce$_3T_4M_{13}$ systems at low temperature, the exact nature of their magnetic ground states remain unresolved. One notable exception is Ce$_3$Ir$_4$Sn$_{13}$, whose long-range antiferromagnetic ordering transition is marked by a sharp lambda-like heat capacity anomaly~\cite{sato1993magnetic,takayanagi1994two}. 

In this paper, we report the discovery of Ce$_3$Ir$_4$Ge$_{13}$ in single crystal form. The crystal structure of Ce$_3$Ir$_4$Ge$_{13}$ is a tetragonally-distorted derivative of the ideal Remeika phase, with three distinct rare earth sites. The transport properties of Ce$_3$Ir$_4$Ge$_{13}$ defy simple categorization: the electrical resistivity is large for an intermetallic compound ($1-2$~m$\Omega$-cm) and nearly temperature independent while Hall coefficient measurements show it is a low-carrier semimetal. Furthermore, Ce$_3$Ir$_4$Ge$_{13}$ is a rare example of a structurally ordered inhomogeneous mixed valence compound, where magnetic Ce$^{3+}$ preferentially occupies two of the three crystallographic rare earth sites while the third is occupied by non-magnetic Ce$^{4+}$. Despite this more dilute magnetic environment, the onset of magnetic correlations in Ce$_3$Ir$_4$Ge$_{13}$ is actually enhanced to higher temperatures in comparison to other Ce$_3T_4M_{13}$ compounds. We find that short range magnetic order sets in at $T_{\text{SRO}} = 10$~K, with a second transition, possibly long-range magnetic order or spin freezing, occurring at $T^*=1.7$~K. The intertwined structural, electronic, and magnetic properties of Ce$_3$Ir$_4$Ge$_{13}$ set it apart, even amongst the known Ce$_3T_4M_{13}$ materials that already show a wide range of ground states.

\section{Methods}

Single crystals of Ce$_3$Ir$_4$Ge$_{13}$ were grown by a self-flux method with a Ce:Ir:Ge ratio of 2:3:15. The starting reagents were combined in an alumina crucible and sealed in an evacuated quartz tube under a partial pressure of argon. The metals were melted and homogenized at 1200~$^{\circ}$C and cooled at 2~$^{\circ}$C/hour to 965~$^{\circ}$C, at which point the crystals were separated from excess liquid flux using a centrifuge. The as-grown crystals of Ce$_3$Ir$_4$Ge$_{13}$ had typical sizes of 1~mm$^3$. We observed a very narrow window of formation for the Ce$_3$Ir$_4$Ge$_{13}$ phase with small changes in starting composition yielding either CeIrGe$_3$ or CeIr$_3$Ge$_7$. The growth conditions described here yielded a mixture of Ce$_3$Ir$_4$Ge$_{13}$ and CeIr$_3$Ge$_7$~\cite{rai2018ceir} crystals, which were distinguished by their characteristic morphologies (\emph{e.g.} see Refs.~\cite{rai2018anomalous,rai2019low}). Attempts to prepare polycrystalline Ce$_3$Ir$_4$Ge$_{13}$ via arc-melting were unsuccessful (majority phase CeIrGe$_3$) suggesting that this material is not congruently melting. All the data presented here, including crystallography, was collected on pieces from one crystal, which was the largest and most well-formed specimen.

Single crystal x-ray diffraction measurements were performed with a Bruker D8 Quest Kappa diffractometer equipped with an I$\mu$S microfocus source (Mo $K\alpha$, $\lambda = 0.71073$~\AA), a HELIOS optics monochromator, and a PHOTON II CPAD detector. The Bruker SAINT program was used to integrate the diffraction data and the absorption correction was performed using the Bruker program SADABS2016/2 (multiscan method)~\cite{krause2015comparison}. Preliminary models were generated using intrinsic phasing methods in SHELXT~\cite{sheldrick2015shelxt}, and anisotropically refined using SHELXL2014~\cite{sheldrick2015crystal}. Powder x-ray diffraction measurements were performed on crushed single crystal using a Bruker D8 Advance with Cu $K\alpha$ radiation and the data were analyzed via Rietveld refinements with TOPAS. 

DC magnetic susceptibility measurements between 0.45 and 400~K were carried out in a Quantum Design (QD) Magnetic Property Measurement System (MPMS) equipped with a $^3$He insert. Heat capacity measurements between 2 and 300~K were performed in a QD Physical Property Measurement System (PPMS) and additional measurements down to 0.1~K were performed in a QD Dynacool equipped with a dilution refrigerator insert. Temperature dependent AC electrical resistivity measurements were performed with current $i \parallel [100] = 2$~mA and frequency $f = 462.02$~Hz in a QD PPMS with a $^3$He insert. To determine the Hall coefficient, we applied a magnetic field perpendicular to the current, $H\perp i \parallel[100]$, and measured the transverse voltage with $H = 9$~T and $-9$~T to cancel out any longitudinal contribution.  


\section{Results and Discussion}


\subsection{Crystal Structure}
\begin{table}[tbp]
\caption{Crystallographic parameters for the single crystal x-ray refinement of Ce$_3$Ir$_4$Ge$_{13}$ at $T = 298$~K.}
\begin{tabular}{p{5.7cm}C{2.7cm}}
\hline
\hline
Formula & Ce$_3$Ir$_4$Ge$_{13}$ \\
\hline
Space group         & $I4_1/amd$   \\
$a$ (\AA) & 18.069(3)    \\
$c$ (\AA) & 18.132(4)    \\
$V$ (\AA$^3$) & 5920(2)   \\
$Z$  & 16           \\
$\theta$ range ($^{\circ}$)    & 3.2-30.3   \\
Absorption coefficient (mm$^{-1}$)   & 70.71         \\
Measured reflections       & 107397       \\
Independent reflections     & 2563         \\
$R_{int}$       & 0.069        \\
$R_1(F)^a$ 
& 0.032        \\
$wR_2(F^2)^b$       & 0.061       \\ 
\hline
\hline
\end{tabular}
$^{a} R_1 = \sum||F_o| - |F_c|| / \sum |F_o|$ \\
$^{b} wR_2 = \{\sum[w({F_o}^2 - {F_c}^2)^2]/ \sum[w({F_o}^2)^2]\}^{1/2}$ 
\label{Table1} 
\end{table}

\begin{table*}[tbp]
\caption{Atomic positions and thermal parameters for Ce$_3$Ir$_4$Ge$_{13}$ in the $I4_1/amd$ space group.}
\begin{tabular}{p{1.3cm}C{2.2cm}C{2.8cm}C{2.8cm}C{2.8cm}C{2.8cm}C{2.2cm}}
\hline
\hline
     & Wyckoff & $x$                      & $y$        & $z$        & $U_{eq}$   & Occ.    \\ \hline
Ce1  & $16h$     & \sfrac{1}{2}    & 0.37559(4) & 0.00001(5) & 0.0056(1)  & 1       \\
Ce2  & $16g$     & 0.25120(3)      & 0.49880(3) & \sfrac{1}{8}        & 0.0064(1)  & 1       \\
Ce3  & $16h$     & 0               & 0.37382(4) & 0.00155(5) & 0.0058(1)  & 1       \\
Ir1  & $32i$     & 0.12515(2)      & 0.37445(2) & 0.12632(2) & 0.00397(9) & 1       \\
Ir2  & $16g$     & 0.37574(2)      & 0.37426(2) & \sfrac{1}{8}        & 0.0040(1)  & 1       \\
Ir3  & $16g$     & 0.12558(2)      & 0.62442(2) & \sfrac{1}{8}        & 0.0039(1)  & 1       \\
Ge1  & $16h$     & 0.24978(9)      & \sfrac{3}{4}        & 0.00434(8) & 0.0097(2)  & 1       \\
Ge2  & $16h$     & 0               & 0.32892(8) & 0.16970(8) & 0.0068(3)  & 1       \\
Ge3  & $32i$     & 0.25065(6)      & 0.32863(5) & 0.16882(6) & 0.0080(2)  & 1       \\
Ge4  & $32i$    & 0.17145(5)      & 0.41729(6) & 0.00068(6) & 0.0073(2)  & 1       \\
Ge5  & $16h$     & 0               & 0.6571(1)  & 0.17497(9) & 0.0102(3)  & 1       \\
Ge6  & $16h$     & 0.12839(9)      & \sfrac{1}{4}        & 0.06546(9) & 0.0094(3)  & 1       \\
Ge7  & $32i$     & 0.31375(6)      & 0.37803(6) & 0.00137(6) & 0.0105(2)  & 1       \\
Ge8  & $32i$     & 0.06938(6)      & 0.50011(6) & 0.11122(6) & 0.0107(2)  & 1       \\
Ge9A & $16h$     & \sfrac{1}{2}    & 0.3362(5)  & 0.1738(3)  & 0.009(1)   & 0.65(2) \\
Ge9B & $16h$     & \sfrac{1}{2}    & 0.3576(6)  & 0.1822(7)  & 0.009(1)   & 0.35(2) \\ \hline
\hline
\end{tabular}
\label{Table2}
\end{table*}

\begin{figure}[tb]
\linespread{1}
\par
\includegraphics[width=3.2in]{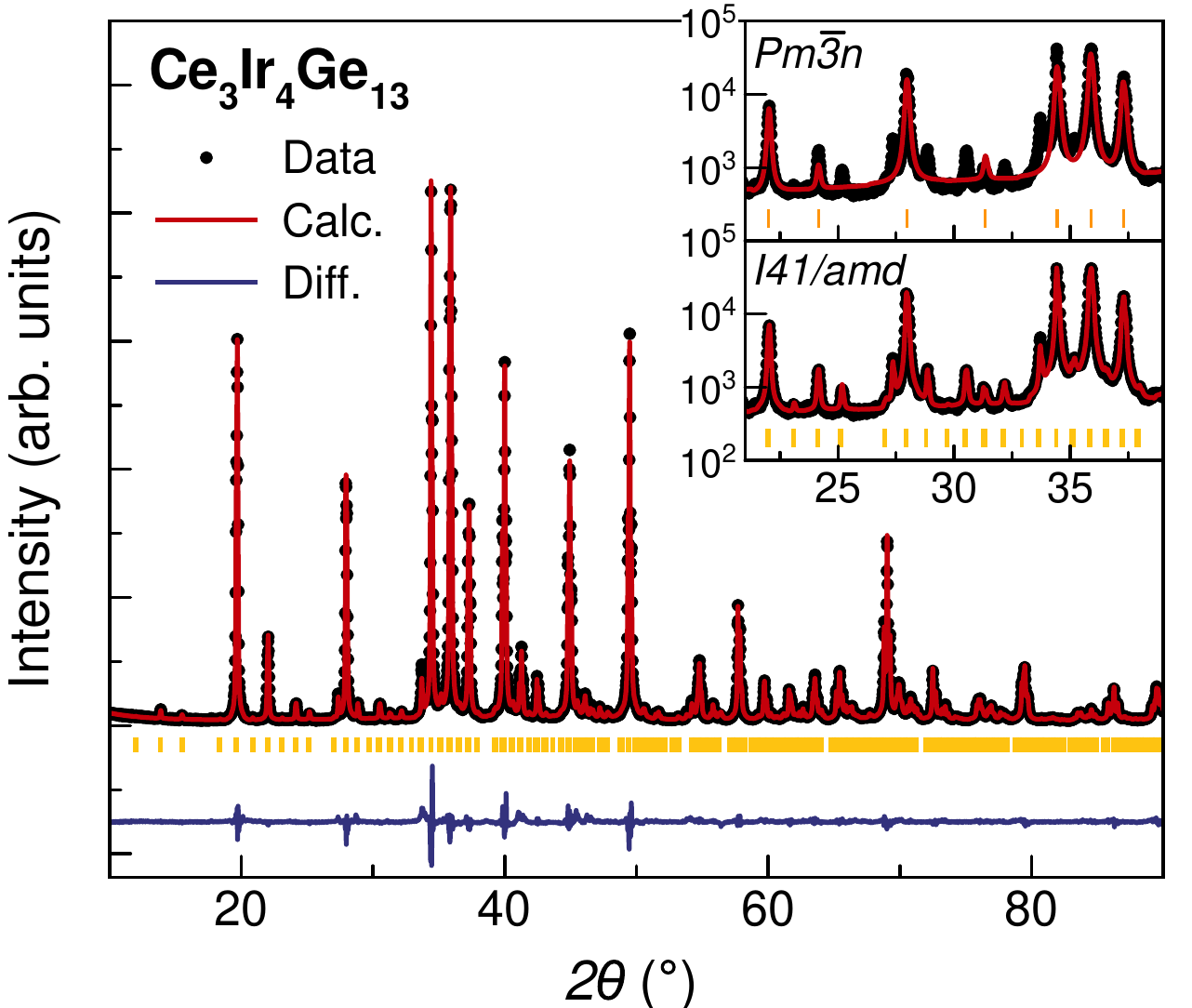}
\par
\caption{Rietveld refinement of the powder x-ray diffraction pattern of Ce$_3$Ir$_4$Ge$_{13}$ measured at room temperature using the $I4_1/amd$ structural model that was derived from single crystal x-ray diffraction. The Bragg peak positions are indicated by the vertical yellow lines. Inset: The undistorted cubic $Pm\bar{3}n$ structure (top) fails to account for many of the observed reflections while the tetragonally distorted $I4_1/amd$ structure (bottom) gives excellent agreement with the data. Note the inset is plotted on a log intensity scale.}
\label{Xray} 
\end{figure}

X-ray diffraction measurements on Ce$_3$Ir$_4$Ge$_{13}$ reveal that this material crystallizes in a Remeika-type phase. Rietveld refinements of our powder x-ray diffraction data with the cubic $Pm\bar{3}n$ space group of the parent compound for this structure type gave poor agreement, failing to account for many of the observed reflections ($R_{wp} = 24.0\%$, top inset of Fig.~\ref{Xray}). The appropriate structural model was determined via single crystal x-ray diffraction measurements, the results of which are summarized in Table~\ref{Table1}. Ce$_3$Ir$_4$Ge$_{13}$ crystallizes in the tetragonally distorted $I4_1/amd$ space group, with lattice parameters $a = 18.069(3)$~\AA~and $c = 18.132(4)$~\AA, giving a unit cell volume that is eight times larger than the undistorted parent space group. Rietveld refinement of the powder data with this structural model yielded excellent agreement as shown in Fig.~\ref{Xray} ($R_{wp} = 7.37\%$). The $I4_1/amd$ tetragonally distorted modification of the Remeika phase has previously been observed for both Lu$_3$Ir$_4$Ge$_{13}$ and Yb$_3$Ir$_4$Ge$_{13}$~\cite{oswald2017proof,rai2019low}. Given that Ce and Lu sit at opposite ends of the lanthanide block, it is very likely that this distorted structure can be stabilized for all members of the $R_3$Ir$_4$Ge$_{13}$ family. 

\begin{figure}[tbp]
\linespread{1}
\par
\includegraphics[width=3.2in]{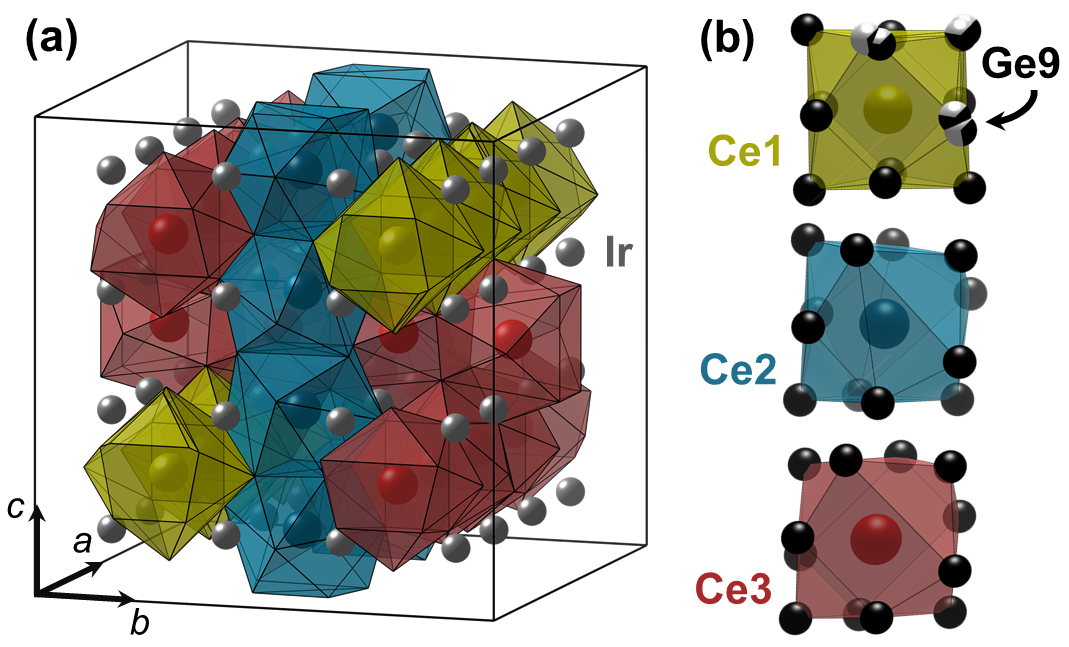}
\par
\caption{(a) The tetragonally distorted $I4_1/amd$ crystal structure of Ce$_3$Ir$_4$Ge$_{13}$ determined from single crystal x-ray diffraction measurements. The three unique cerium sites are indicated by the yellow (Ce1), blue (Ce2), and red (Ce3) polyhedra, each of which is a distorted CeGe$_{12}$ icosahedron. The Ir atoms are given by the grey spheres. (b) The germanium atoms (black spheres) sit at the vertices of these polyhedra and are omitted from (a) for visual clarity. The positionally disordered Ge9 site is coordinated with only Ce1.}
\label{Structure} 
\end{figure}

In the undistorted $R_3T_4M_{13}$ structure with space group $Pm\bar{3}n$, $R$ and $T$ each occupy a single unique crystallographic site while $M$ is distributed over two sites, with only two adjustable atomic coordinates for the whole structure. The $I4_1/amd$ tetragonal distortion adds a significant number of lattice degrees of freedom. $R$ and $T$ are each split over three crystallographic sites while $M$ is split over nine sites, and the number of adjustable atomic coordinates catapults to 32. The atomic positions and thermal parameters for Ce$_3$Ir$_4$Ge$_{13}$ determined from refinement of the single crystal x-ray diffraction data are provided in Table~\ref{Table2}. This refinement reveals that there is no site-mixing or vacancies for either the Ce or Ir sublattices. There is, however, positional disorder associated with the Ge9 site, which is split into sites Ge9A and Ge9B with 65\% and 35\% occupancies, respectively. This is similar to what was previously observed for Lu$_3$Ir$_4$Ge$_{13}$, but interestingly, in that compound it was the Ge1 site that was split instead of Ge9. The crystal structure of Ce$_3$Ir$_4$Ge$_{13}$ is presented in Fig.~\ref{Structure}, where the yellow, red, and blue polyhedra are the distorted CeGe$_{12}$ icosahedra for the three unique Ce sites. The disordered Ge9 site makes up three of the nearest neighbors for the Ce1 site but does not appear in the local environments of either Ce2 or Ce3.


\subsection{Electronic Properties}

\begin{figure}[tbp]
\linespread{1}
\par
\includegraphics[width=3.2in]{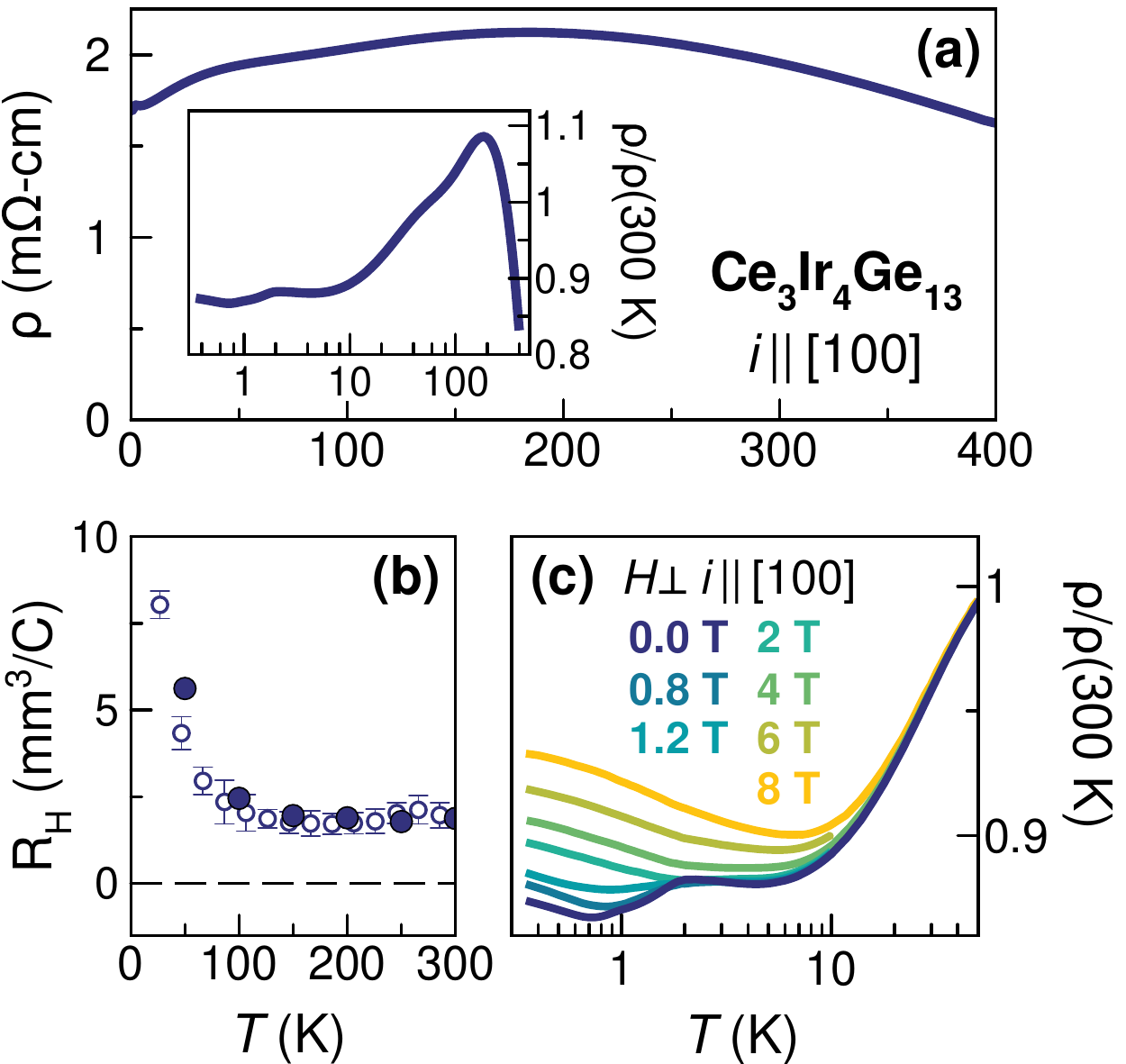}
\par
\caption{(a) The electrical resistivity of Ce$_3$Ir$_4$Ge$_{13}$ measured with $i \parallel [100]$ showing distinctly non-metallic behavior. The inset shows the same resistivity data for Ce$_3$Ir$_4$Ge$_{13}$ normalized to its room temperature value on a contracted $y$-axis and a logarithmic temperature scale. (b) Temperature dependent Hall coefficient where the open symbols are measured by sweeping temperature at $H = 9$~T and the filled symbols were measured by field sweeps at constant temperature. (c) The field dependence of the low temperature resistivity of Ce$_3$Ir$_4$Ge$_{13}$ measured with $H\perp i \parallel [100]$.}
\label{Transport} 
\end{figure}

The temperature dependent electrical resistivity of Ce$_3$Ir$_4$Ge$_{13}$ is presented in Fig.~\ref{Transport}(a). Over the full measured temperature range, from 0.4 K to 400~K, the resistivity of Ce$_3$Ir$_4$Ge$_{13}$ is large (around 2 m$\Omega$-cm) and nearly temperature independent. The exact magnitude of the resistivity was observed to vary in different samples ($\rho(300~\text{K}) = 1-2$~m$\Omega$-cm), likely an artifact of geometric factors associated with the small size of our crystals ($< 1$~mm$^3$); however, the qualitative features of the resistivity are sample independent. There is a broad peak  centered near 180 K, as can be seen more clearly in the inset of Fig.~\ref{Transport}(a). On cooling, a small drop in the resistivity is observed around $T^*=1.7$~K, which, as will be shown below, is likely due to the loss of spin-disorder scattering at the magnetic ordering or spin freezing transition. 

Non-metallic transport behavior has been observed in a large number of $R_3T_4M_{13}$ compounds. In particular, in both $R_3$Ir$_4$Ge$_{13}$ ($R=$~Lu and Yb) the resistivity monotonically increases with decreasing temperature (\emph{i.e.} d$\rho$/d$T < 0$) before the Lu analog enters a superconducting state at $T_C = 1.4$ K and Kondo scattering yields a diverging resistivity in the Yb analog~\cite{rai2019low}. Despite this non-metallic resistivity behavior, optical conductivity and Hall coefficient measurements have confirmed that both are semimetals with low carrier concentrations. Hall coefficient measurements on Ce$_3$Ir$_4$Ge$_{13}$ presented in Fig.~\ref{Transport}(b) reveal hole-like carriers at all measured temperatures. The magnitude of $R_H$ is a factor of two or more larger at all temperatures than that of the $R~=$ Lu and Yb analogs, suggesting an even smaller concentration of charge carriers, rendering Ce$_3$Ir$_4$Ge$_{13}$ a low-carrier semimetal. 


Finally, we consider the effect of a magnetic field on the electrical resistivity of Ce$_3$Ir$_4$Ge$_{13}$, as shown in Fig.~\ref{Transport}(c). The resistivity above the anomaly at $T^*=1.7$~K is field-independent up to $H = 2$~T. With increasing field, the lowest temperature resistivity monotonically increases. At the highest measured field, $H = 8$~T the resistivity is increased relative to the zero field value up to approximately 20 K and field-independent at higher temperatures. Below 0.7 K, the zero field data shows a small upturn which extends over almost a decade in temperature when the field is increased to $H = 8$~T. While the divergent resistivity is reminiscent of the Kondo effect, there is little evidence from specific heat data (below) to support this scenario in Ce$_3$Ir$_4$Ge$_{13}$.


\subsection{Magnetic Properties}

\begin{figure}[tbp]
\linespread{1}
\par
\includegraphics[width=3.2in]{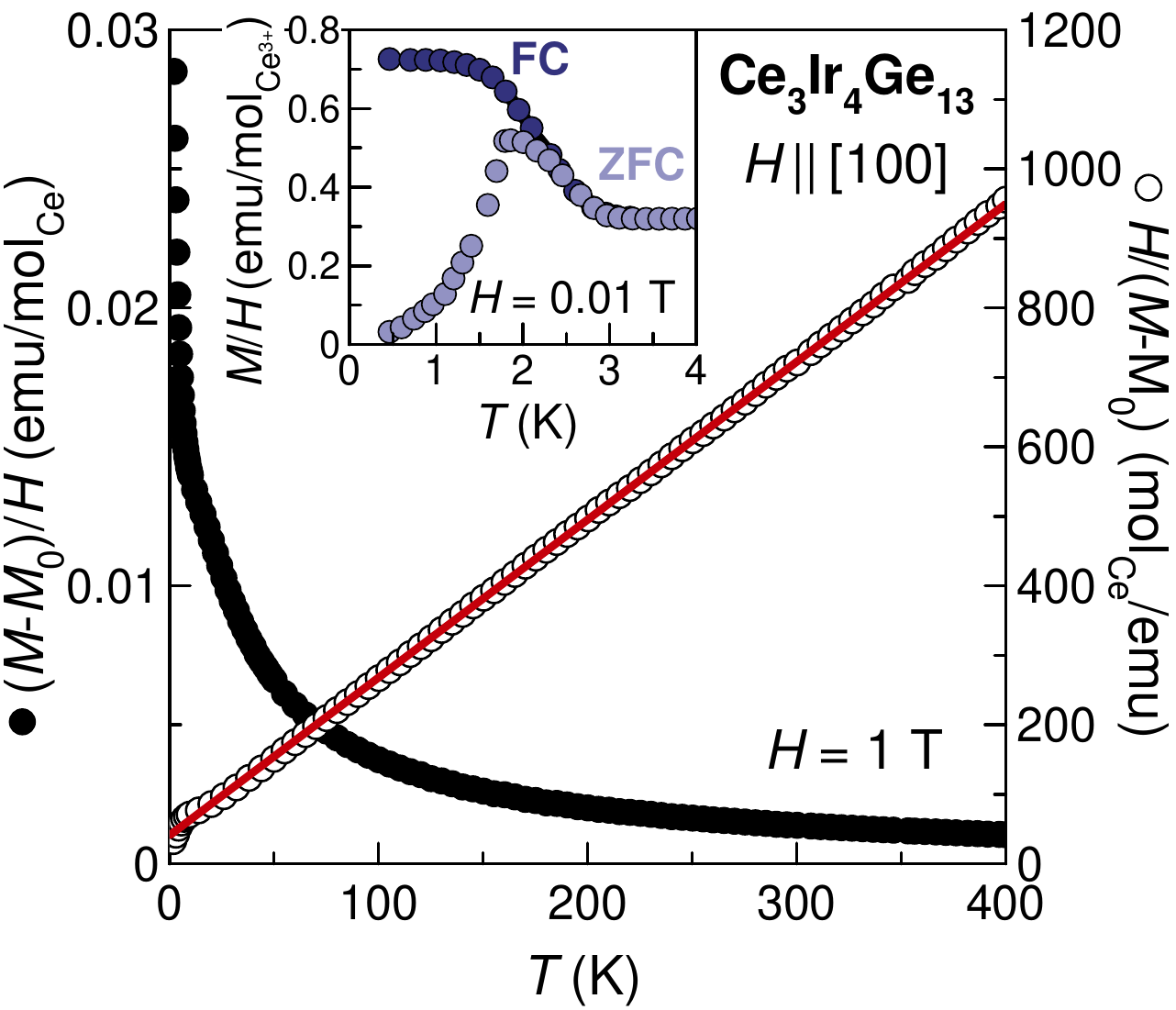}
\par
\caption{The magnetic susceptibility (left axis) and inverse magnetic susceptibility (right axis) of Ce$_3$Ir$_4$Ge$_{13}$ measured in an $H \parallel [100] = 1$~T field. The red line is a fit to the Curie-Weiss law between 50 and 400~K, giving $\theta_{CW} = -17.3(3)$~K and an effective moment of $\mu^{\text{exp}}_{\text{eff}} = 1.87~\mu_B$/Ce. Inset: the field-cooled (FC) and zero-field-cooled (ZFC) low temperature susceptibility measured in a field of $H = 0.01$~T revealing a magnetic transition at $T^* = 1.7$~K.}
\label{Chi} 
\end{figure}

The DC magnetic susceptibility of Ce$_3$Ir$_4$Ge$_{13}$ with $H \parallel [100] = 1$~T is shown as filled symbols in Fig.~\ref{Chi}, where a small temperature-independent contribution $M_0$ has been subtracted. The inverse susceptibility (open symbols, right axis) is linear over a wide temperature range consistent with the Curie-Weiss behavior expected for a local $4f$ moment. A fit to the Curie-Weiss equation between 50 and 400 K, indicated by the solid line, gives $\theta_{CW} = -17.3(3)$ K and an effective moment of $\mu^{\text{exp}}_{\text{eff}} = 1.87~\mu_B$/Ce. This effective moment value falls significantly short of the expected value $\mu^{\text{calc}}_{\text{eff}} = 2.54~\mu_B$/Ce for Ce$^{3+}$. A scenario of an intermediate valence (\emph{i.e.} Ce$^{3+\delta}$ where $\delta$ is temperature-dependent) can be excluded as this would result in a broad hump in the magnetic susceptibility~\cite{sales1975susceptibility,rai2016intermediate}, which is inconsistent with our data. The most likely explanation for this reduced moment is that Ce$_3$Ir$_4$Ge$_{13}$ contains an inhomogeneous mixture of magnetic Ce$^{3+}$ and non-magnetic Ce$^{4+}$ ions. In a scenario of preferential occupancy of the three inequivalent Ce sites, our susceptibility data is most consistent with two Ce$^{3+}$ sites and one Ce$^{4+}$ site. The experimental effective moment (scaled per Ce$^{3+}$) becomes $\mu^{\text{exp}}_{\text{eff}} = 2.30~\mu_B$/Ce$^{3+}$, very close to the theoretical value. Although our data does not allow us to identify which of the Ce sites is non-magnetic, one possibility is that the splitting of Ge9 induces a Ce$^{4+}$ valence at the Ce1 site. Furthermore, the nearest neighbor Ce-Ge distances are on average, slightly smaller than Ce2 and Ce3, which may be consistent with the expected smaller ionic radius of Ce$^{4+}$. However, these Ce-Ge contacts are non-bonding and therefore making a definitive assessment will require further measurements. For the remainder of the paper we normalize our data to two Ce$^{3+}$ per formula unit. 

We next turn to the low temperature DC magnetic susceptibility measured with $H = 0.01$~T, which is shown in the inset of Fig.~\ref{Chi}. A sharp increase in the susceptibility and a splitting of the zero-field-cooled (ZFC) and field-cooled (FC) measurements marks a likely magnetic ordering transition or spin freezing transition at $T^* = 1.7$~K. This hysteresis is largely closed by a field of 0.1 T and completely closed by 1~T (not shown). As shown in the main panel in Fig. \ref{Chi}, the susceptibility deviates from Curie-Weiss behavior at a slightly higher temperature, approximately 10 K, which is possibly associated to the formation of short range correlations. The heat capacity data, presented next, lends credence to such a scenario.

\begin{figure}[tbp]
\linespread{1}
\par
\includegraphics[width=3.2in]{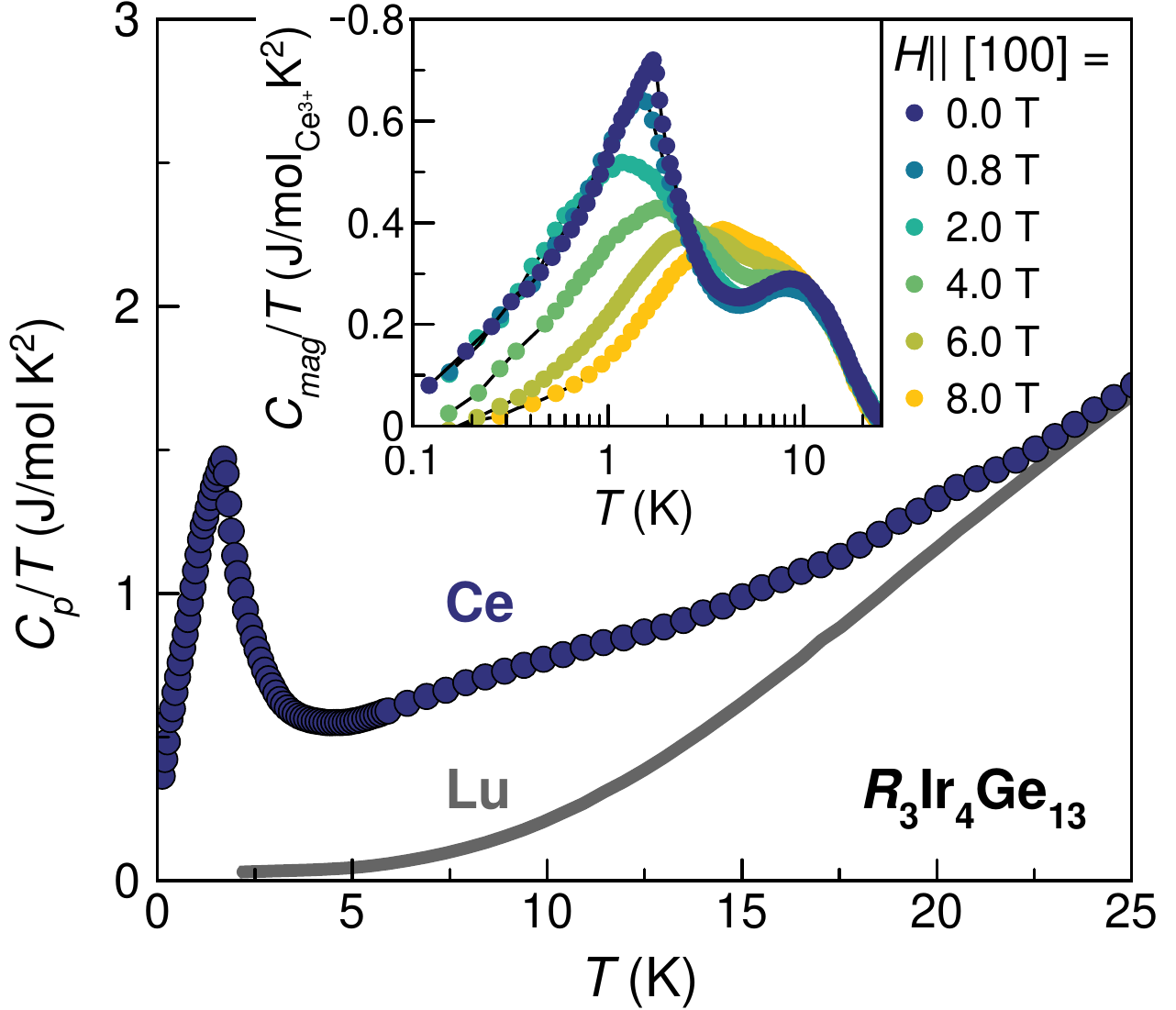}
\par
\caption{Temperature dependence of the low temperature heat capacity, $C_p/T$, for Ce$_3$Ir$_4$Ge$_{13}$ and the heat capacity of its non-magnetic lattice analog Lu$_3$Ir$_4$Ge$_{13}$ scaled to correct for the mass difference between Lu and Ce. The data for Lu$_3$Ir$_4$Ge$_{13}$ is truncated just above its superconducting transition at $T_C = 1.4$~K. The non-magnetic contribution was extrapolated to the lowest temperatures by fitting the data for Lu$_3$Ir$_4$Ge$_{13}$ to $\gamma T + \beta T^3$. The inset shows the magnetic heat capacity, $C_{mag}/T$, in varying applied fields with $H \parallel [100]$.}
\label{Cp} 
\end{figure}

\begin{figure}[tbp]
\linespread{1}
\par
\includegraphics[width=3.2in]{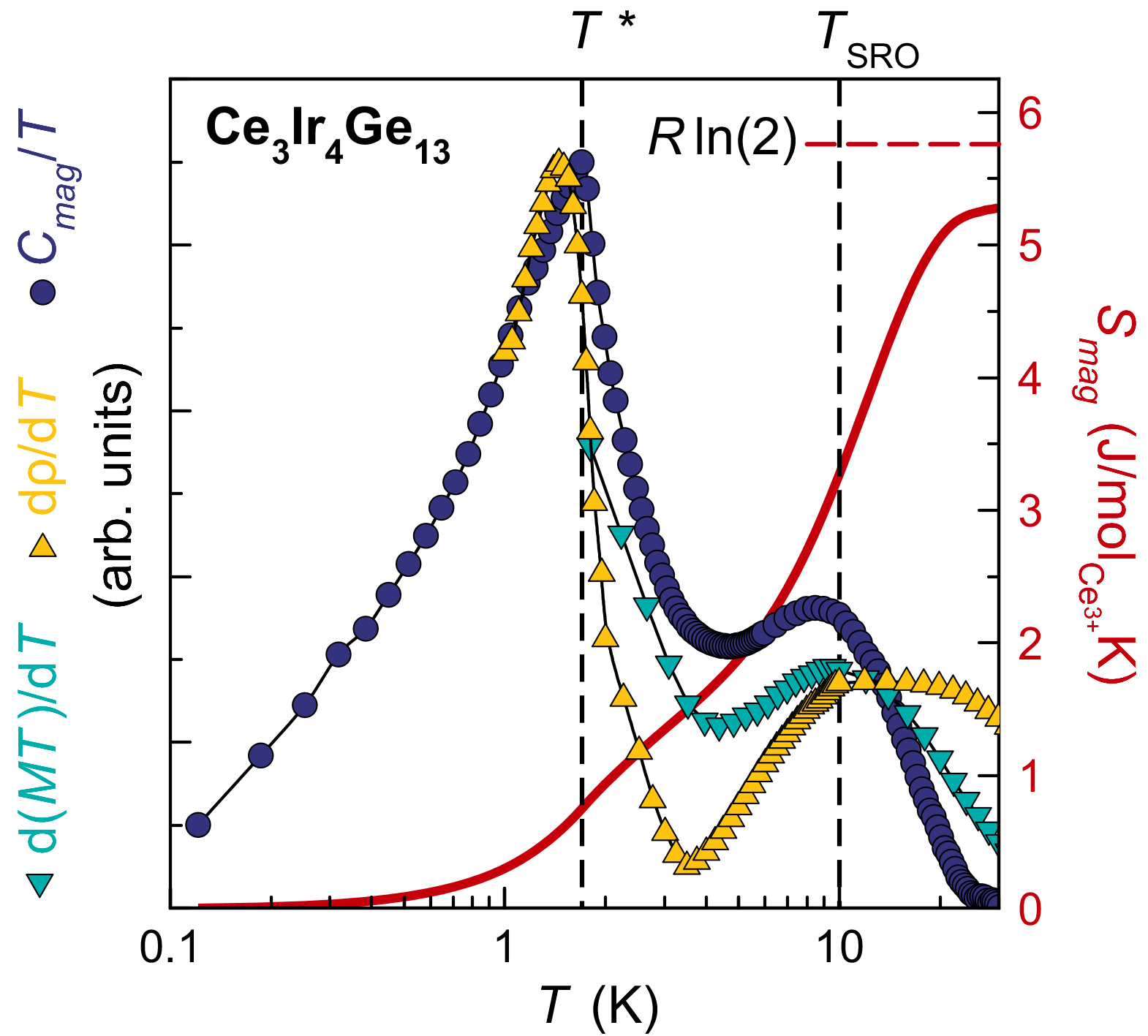}
\par
\caption{Bulk property measurements on Ce$_3$Ir$_4$Ge$_{13}$, d$(MT)$/d$T$ (teal triangles), d$\rho$/d$T$ (yellow triangles), and $C_{mag}/T$ (purple circles), reveal that the magnetic state evolves at two distinct temperature scales, $T_{\text{SRO}} = 10$~K, which we interpret as the onset of short-range magnetic correlations and $T^* = 1.7$~K, which may be a long range ordering or spin freezing transition. The entropy release associated with the two heat capacity anomalies is close to the expected $R\ln{2}$ per Ce$^{3+}$ for a crystal field ground state doublet.}
\label{Derivatives} 
\end{figure}

The zero field heat capacity $C_p/T$ of Ce$_3$Ir$_4$Ge$_{13}$ (symbols) and that of its non-magnetic analogue Lu$_3$Ir$_4$Ge$_{13}$ (solid line) scaled by temperature are presented in Fig.~\ref{Cp}. To account for the mass difference between these two materials, we have scaled the heat capacity of Lu$_3$Ir$_4$Ge$_{13}$ by $\mu = \sqrt{\sfrac{M_{\text{Ce}}}{M_{\text{Lu}}}}$ where $M_{\text{Ce}}$ and $M_{\text{Lu}}$ are the molecular weights of Ce$_3$Ir$_4$Ge$_{13}$ and Lu$_3$Ir$_4$Ge$_{13}$, respectively. Two prominent features are present in the Ce$_3$Ir$_4$Ge$_{13}$ data but not for Lu$_3$Ir$_4$Ge$_{13}$, and therefore must be magnetic in origin: a peak centered at $T^*=1.7$~K and a much broader hump centered around 10~K. The higher temperature anomaly, which we label as $T_{\text{SRO}}$, can be seen more clearly in the magnetic heat capacity $C_{mag}/T$ shown in the inset of Fig.~\ref{Cp}, which is isolated by subtracting the scaled data for Lu$_3$Ir$_4$Ge$_{13}$. The broad hump at $T_{\text{SRO}}=10$~K is field independent and cannot be fit to the characteristic form of a Schottky anomaly, ruling out a low-lying crystal electric field level as its origin. The sharper peak at $T^* = 1.7$~K has a weak field-dependence: it first moves down in field for $H \leq 2$~T, and then it broadens and ultimately merges with the higher temperature hump by $H = 8$~T. The combined entropy $S_{mag}$ of these two magnetic anomalies (line, right axis of Fig.~\ref{Derivatives}) is close to $R\ln{2}$ when normalized to the two Ce$^{3+}$ per formula unit determined by the analysis of the magnetic susceptibility data. This would suggest a doublet ground state and rule out strong Kondo screening, as may have been suggested by the diverging resistivity (Fig.~\ref{Transport}(c)). 

Given its lack of field dependence, it is tempting to invoke a non-magnetic origin for the broad heat capacity anomaly at $T_{\text{SRO}}=10$~K, akin to the quadrupolar order observed in Ce$_3$Pd$_{20}$Ge$_6$~\cite{kitagawa1996possible}. However, unlike Ce$_3$Pd$_{20}$Ge$_6$, which has a ground state quartet, the crystal field ground state in Ce$_3$Ir$_4$Ge$_{13}$ is a Kramers doublet, as expected from the low point group symmetry at the Ce site and verified by its $R\ln{2}$ entropy release. Thus, Ce$^{3+}$ in Ce$_3$Ir$_4$Ge$_{13}$ cannot have quadrupolar degrees of freedom. While this higher temperature anomaly is the most pronounced in heat capacity data, it is correlated with features in both the magnetic susceptibility and resistivity data. Fig.~\ref{Derivatives} shows $C_{mag}/T$ (circles) alongside the derivatives d$(MT)$/d$T$ (down triangles) and d$\rho$/d$T$ (up triangles), since the three curves are expected to feature maxima at the same temperatures, according to Fisher's relation for magnetic metals~\cite{fisher1962relation,fisher1968resistive}. Therefore, the anomaly at $T_{\text{SRO}}=10$~K is most likely due to short-range magnetic correlations. 

\section{Discussion and Summary}

Despite their crystallographic similarities, the magnetic and electronic properties of Ce$_3$Ir$_4$Ge$_{13}$ are strikingly different from those of other Ce-based Remeika compounds, but also from those of the Yb analogue Yb$_3$Ir$_4$Ge$_{13}$. While both compounds are low-carrier semimetals, Yb$_3$Ir$_4$Ge$_{13}$ has a strong Kondo effect~\cite{rai2019low}, which is not evident in Ce$_3$Ir$_4$Ge$_{13}$. Magnetic susceptibility measurements on Yb$_3$Ir$_4$Ge$_{13}$ are consistent with Yb$^{3+}$ valence only, and not the inhomogeneous mixed valence we observe in the case of Ce$_3$Ir$_4$Ge$_{13}$. Both compounds have features in their bulk property measurements suggestive of a magnetic ordering transition (at $T^*=0.9$~K in the case of $R=$~Yb and $T^*=1.7$~K for $R =$ Ce). However, attempts to find definitive signatures of long range order in Yb$_3$Ir$_4$Ge$_{13}$ with neutron scattering and muon spin relaxation experiments have not been successful~\cite{rai2019low,iwasa2018magnetic}. Similar measurements on Ce$_3$Ir$_4$Ge$_{13}$ are planned.

We conclude by placing Ce$_3$Ir$_4$Ge$_{13}$ in context with respect to its broader Ce$_3T_4M_{13}$ family. While several of these materials have been found to form in various structurally distorted versions of the ideal $Pm\bar{3}n$ Remeika phase, Ce$_3$Ir$_4$Ge$_{13}$ is the only one known to crystallize in the $I4_1/amd$ structure. This tetragonal distortion generates three unique Ce sites, which appear to be an essential ingredient to its inhomogeneous mixed valence character. It is interesting to note that a completely different origin for mixed valence has been previously observed in both Ce$_3$Ru$_4$Ge$_{13}$~\cite{ghosh1995crystal} and Ce$_3$Os$_4$Ge$_{13}$~\cite{prakash2016ferromagnetic}. These compounds have significant site-mixing and it is only the disordered Ce atoms occupying the Ge sites that carry a magnetic moment while the majority of Ce atoms are non-magnetic. In contrast, our single crystal x-ray diffraction measurements reveal that Ce$_3$Ir$_4$Ge$_{13}$ is fully site-ordered and the only structural disorder relates to a splitting of the Ge9 site. Magnetic property measurements on these two site-mixed inhomogeneous mixed valence compounds, Ce$_3$Ru$_4$Ge$_{13}$ and Ce$_3$Os$_4$Ge$_{13}$, are very similar to one another, albeit the interpretations offered are different: the properties of the former were attributed to spin glass behavior~\cite{ghosh1995crystal} while the latter is claimed to show ferromagnetic order~\cite{prakash2016ferromagnetic}.

The magnetic and electronic properties of Ce$_3$Ir$_4$Ge$_{13}$ most closely resemble those of Ce$_3$Rh$_4$Sn$_{13}$, which has a pure Ce$^{3+}$ valence. Ce$_3$Rh$_4$Sn$_{13}$ has a nearly temperature independent electrical resistivity and a broad heat capacity anomaly centered at 1~K, whose position is field independent up to $H = 2$~T and above which shifts to higher temperatures~\cite{niepmann2001structure,kohler2007low,slebarski2012electronic}, remarkably similar to what we observe for the heat capacity peak at $T^* = 1.7$~K in Ce$_3$Ir$_4$Ge$_{13}$. One significant difference between these two compounds is that there is no higher temperature magnetic anomaly in Ce$_3$Rh$_4$Sn$_{13}$ while we observe a large heat capacity anomaly at $T_{\text{SRO}}=10$~K in Ce$_3$Ir$_4$Ge$_{13}$. While the ensemble of properties we observe at $T^* = 1.7$~K in Ce$_3$Ir$_4$Ge$_{13}$ make it tempting to assign this as a long-range ordering transition, it has borne out that the same set of signatures in Ce$_3$Rh$_4$Sn$_{13}$ do not indicate long-range order based on neutron diffraction measurements~\cite{suyama2018chiral}. Inelastic neutron scattering measurements on Ce$_3$Rh$_4$Sn$_{13}$ have uncovered the formation of a low energy, $E = 0.2$~meV, mode in Ce$_3$Rh$_4$Sn$_{13}$ proposed to originate from an exchange splitting of the crystal field ground state doublet~\cite{adroja2008crystal}. Similar neutron scattering measurements on Ce$_3$Ir$_4$Ge$_{13}$ will be of the utmost interest as will frequency dependent AC susceptibility measurements to search for signatures of glassiness.


\begin{acknowledgments}
\section{Acknowledgments}
AMH acknowledges support from the Rice Center for Quantum Materials and the Natural Sciences and Engineering Research Council (NSERC) of Canada.  Work at Rice was supported in part by the Gordon and Betty Moore Foundation EPiQS Initiative through Grant No. GBMF 4417 and US DOE BES DE-SC0019503. Work at UT Dallas was supported by NSF DMR-1700030. Work at McMaster was supported by NSERC and the Canadian Foundation for Innovation.

\end{acknowledgments}

\bibliography{References}

\end{document}